\documentclass{elsart}
\usepackage{amssymb}
\usepackage{harvard}
\input psfig.sty

\newcommand{\degs}{\mbox{\(^\circ \)}}
\newcommand{\gr}{$\gamma$-ray \, }
\newcommand{\grs}{$\gamma$-rays \,}

\begin{document}
\begin{frontmatter}
\title{Exploring physics  of rotation powered pulsars
with sub-10 GeV imaging atmospheric Cherenkov telescopes}

\author[mpik]{F.A. Aharonian\thanksref{1}} 
\author[mepi]{S.V.Bogovalov\thanksref{2}}

\thanks[1]{E-mail: Felix.Aharonian@mpi-hd.mpg.de}  
\thanks[2]{E-mail: bogoval@axpk40.mephi.ru}

\address[mpik]{MPI f\"ur Kernphysik, Sauperfercheckweg 1,
D-69117 Heidelberg, Germany}
\address[mepi]{Astrophysics Institute at the
Moscow Engineering Physics Institute,
Kashirskoje Shosse 31, Moscow, 115409, Russia}

\begin{abstract}
We discuss the potential of future sub-10 GeV threshold 
imaging atmospheric Cherenkov telescope arrays for exploring
the physics of rotation  powered pulsars and their interactions
with the  ambient medium  through
relativistic winds and termination shocks.
One such telescope is the high-altitude concept called ``5@5''
recently suggested by  Aharonian et al. (2001).
5@5,  with its enormous
detection area exceeding  $10^{4} \, \rm m^2$ at the threshold
energy of about 5 GeV,   combines two  distinct features  of the
current satellite-borne (large photon  fluxes at GeV  energies)
and ground-based  (large detection areas at TeV energies)
gamma-ray astronomies.  Such an  instrument  would allow 
comprehensive studies of  temporal and spectral  characteristics of  
$\gamma$-ray pulsars in the crucial   5 to 30 GeV energy interval.
An equally  important topic   in the program of pulsar studies by
5@5  would   be  the search for GeV $\gamma$-radiation from other
radio pulsars at a few mVela level.   And finally, the
searches  for pulsed  radiation components in the spectra of a
large fraction of   unidentified EGRET  sources  (suspected to be
pulsars)  without invoking information from  lower  (radio, optical, X-ray)   
frequency domains,  seems to be another  important issue,  because
the periodic signals at  lower energies  could be significantly  suppressed in  many  cases. The  detection rate of $\gamma$-rays
from   ``standard''  EGRET sources by 5@5 is expected to exceed
several  events per one second.  This should  provide an adequate
photon statistics for the search  for periodic signals at the flux level 
of 3 mVela within the observation time of 3 h or so  (a time resolution below
 which any change of a signal's phase can be ignored). The spectral coverage by 5@5
and its  flux sensitivity are  nicely suited  for 
studying  other  aspects
of  pulsar physics and astrophysics, in particular for detecting
 unshocked relativistic pulsar winds,  as well as  for 
 quantifying characteristics of  pulsar driven synchrotron  nebulae through the
inverse   Compton radiation   at energies  between  several GeV
and several 100 GeV.  The Vela pulsar, the brightest $\gamma$-ray
source on the sky,  is  an ideal laboratory for  practical
realization of  these  unique observational  possibilities.
\end{abstract}
\end{frontmatter}
\section{Introduction}

The number of cataloged  radio pulsars  -- single neutron stars
powered by fast  rotation -- exceeds 1200 (Lorimer 2001). Only
seven of them (with two other possible candidates) are reported by
the EGRET team as GeV $\gamma$-ray sources (Thompson 1999, Kaspi
et al. 2000) . Actually, this is not a big surprise, and can be 
explained by   modest flux sensitivities of
$\gamma$-ray instruments. For example, the minimum detectable flux
by EGRET of about $10^{-11} \, \rm erg/cm^2 s$ is  5 orders
of magnitude larger than the sensitivity  achieved by radio
telescopes at GHz frequencies.  This striking difference in
sensitivities is compensated, to a certain extent,   by much
larger energy fluxes  in $\gamma$-rays. While the  pulsars emit in 
the radio band a tiny $\leq 10^{-6}$ fraction of their rotational energy
(Manchester \& Taylor 1977), the $\gamma$-ray  luminosities   of some
of the  EGRET pulsars above $100 \, \rm MeV$ exceed one per cent
of their  spin-down luminosities. It is expected that  the
Gamma-ray Large Area Space telescope (GLAST),   the future major
satellite-based high energy $\gamma$-ray detector with flux
sensitivity as good as $10^{-12} \, \rm erg/cm^2 s$ (Gehrels \&
Michelson 1999),   will  increase the number of {\it radio} pulsars
seen also in  $\gamma$-rays  by at least an order of magnitude  (Thompson
2001). Moreover,   the radio pulsars, which are observable  
only when their  radio beams  cross  the Earth orbit,  are only  a part of
a larger source  population called Rotation Powered  Pulsars
(RPPs, for short).  The \gr beams  are believed  to be
significantly wider than the radio beams and   they 
should not necessarily  overlap with  each other (Yadigaroglu \& Romani 1995). If so,  the RPPs would have improved chances, in principle,  to be detected, in
high energy $\gamma$-rays rather than  at radio wavelengths. This
could be the case of some of the  unidentified EGRET sources, a
large fraction of which  is believed to be radio-quiet pulsars
(see e.g. Grenier 2000).  The  unambiguous and
straightforward   proof of this hypothesis would be the discovery
of $\gamma$-ray pulsations  from these objects. In this regard, it
is crucial to search for the  pulsed $\gamma$-ray components  without
relying on observations at other  energy bands.  Although GLAST
will  be able to perform  effective  searches  for strong unidentified
$\gamma$-ray sources (Thompson 2001),  its  potential in this
regard  will be  still limited, at least  for marginally detected
weak  sources which  most likely  will appear in the  GLAST
source catalog (future ``unidentified   GLAST sources''). An
effective realization  of this  important approach which relies
only on $\gamma$-ray astronomical observations requires large
photon statistics, and therefore much larger detector areas. The
current and forthcoming 
atmospheric imaging Cherenkov telescopes do provide huge (as large
as $10^5 \, \rm m^2$) detection areas. But most of these telescopes
 operate in the energy region
$\geq 100 \, \rm GeV$ which  lies, most probably,  well  beyond the
cutoffs expected in the pulsar spectra.  In contrast, the
recently suggested concept ``5@5''  by Aharonian et al. (2001)  
 --5 GeV energy threshold array of imaging atmospheric Cherenkov
telescopes at 5 km altitude--  which combines the most important
features of the current  space-based (large fluxes at GeV
energies) and ground-based  (large detection areas at TeV
energies)  astronomies, could serve as an ideal tool for pulsar
studies. Remarkably, the analysis of the EGRET pulsars shows that
with  age of pulsars  the fraction of the spin-down luminosity
converted into $\gamma$-rays increases,  and the peak in  the
spectral energy distribution $\nu S_\nu$ shifts towards 10 GeV
(Nel et al. 1996).  If this tendency extends to fainter pulsars,
the chances for detection of many pulsars by 5@5 would be
increased dramatically.

The potential of the 5@5 concept is not limited
to the discovery of  faint  $\gamma$-ray pulsars.
 In addition, 5@5 can  provide  detailed spectroscopy
in one of the key or, perhaps even the most informative, energy regions
 of above several GeV for pulsar physics.
 It is worth noting that  the
EGRET data indicate  that some important changes might 
take place at multi-GeV energies. In particular,
the light curves of all EGRET  pulsars  at energies above
5 GeV  seem to be  different  than at lower  energies (Thompson 2001).
Detailed temporal  and spectroscopic measurements
in this transition region  by 5@5 will hopefully remove
many uncertainties we presently face in the physics of
pulsar magnetospheres.  Accurate measurements of the energy
spectrum of strong  EGRET pulsars in the cutoff region expected
between 5 and 20 GeV would allow, in particular,   to distinguish
between  two currently popular models  of high energy
$\gamma$-radiation  ---  {\it polar cap}  (Daugherty \& Harding
1982, 1996;  Usov \& Melrose 1995)  and {\it outer gap}
(Cheng et al. 1986;   Romani 1996;  Hirotani \&  Shibata 1999)  models.
These models generally predict different fractions
 in the population of radio-quiet versus  radio-loud $\gamma$-ray
pulsars, as well as different spectral characteristics, especially at
 the multi-GeV energy range.  These differences are quite significant,
even with  large uncertainties in  model  parameters
(Harding 2001).
 Thus  5@5 should allow distinction among these models,
  or perhaps even to challenge most scenarios.
Remarkably, this intermediate energy region   can be effectively
studied by both GLAST and 5@5--- a combination of which would provide
 high reliability of results.  
 On one hand  GLAST will  extend the
studies to low energies, down to $\sim 20 \, \rm MeV$,
 while 5@5 can perfectly cover the energy region well above 10 GeV
where GLAST most probably would run out of photons. 
 Presently several ground-based telescopes such as MAGIC, CELESTE and
 STACEE  are pushing  their energy thresholds  below 100 GeV  
 (e.g.  Buckley et al.  2001; Krennrich 2001),
however  none of these is optimized for energies in the crucial 
 10 GeV range. 

The flux sensitivity,  angular resolution, and the spectral
coverage  of 5@5 perfectly match other important aspects
of the pulsar physics and  astrophysics. 
 In particular,  the
search for inverse Compton  $\gamma$-radiation from
unshocked  ultrarelativistic  winds caused by  illumination
of the wind by the IR to X-ray radiation emitted by the
pulsar/neutron star itself (Bogovalov \& Aharonian 2000) or,
in the case of binary  pulsar systems,  
 with illumination 
 provided by the companion star  (Ball \& Kirk 2000), would be an
important topic in the program of pulsar studies by 5@5. 
Because  this radiation is produced  in deep Klein-Nishina regime,
it has a rather specific shape -   unusually narrow   spectra
peaked at   $E_\gamma \sim  \Gamma_{\rm w} m_{\rm e}c^2$,
where   $\Gamma_{\rm w}$ is the wind Lorentz factor.
The detection of this radiation  would be the first
{\it observational} test for  existence of  ultrarelativistic pulsar winds,
and would give unique information  about  the Lorentz  factor
and the site of  creation of the  kinetic energy dominated (KED)
wind.  The large dynamical energy range of 5@5, extending
from several GeV to  several  hundred GeV,  would allow such
a search  for pulsar winds in  the most relevant range of Lorentz
factors  between  $10^4$ to $10^6$.

Another, more traditional,  objective for 5@5
would  be  the detection of inverse Compton $\gamma$-rays
from  synchrotron  nebulae  surrounding pulsars - the  
regions powered by the wind  termination shocks (plerions). 
The TeV radiation detected from  the Crab Nebula by many
groups (for review  see e.g.  Catanese \& Weekes  1999)
 agrees reasonably well with  model calculations  (De Jager \& Harding 1992;
De Jager et al. 1996; Atoyan \& Aharonian 1996;
Aharonian \& Atoyan 1998; Hillas et al. 1998)
performed within the  MHD model of Kennel \& Coroniti (1984).
 According to this model,  the wind is terminated by a standing reverse shock at a distance of about r = 0.1 pc, the shock in turn accelerates
electrons up to energies exceeding  $10^{15} \, \rm eV$ and randomizes their
pitch angles.  The inverse Compton $\gamma$-radiation at GeV/TeV
energies,  combined with  synchrotron optical and  X-ray emission,
contains  information about the  relativistic  electrons and the
nebular magnetic  fields in the  downstream  region of the shock.
 But the expected $\gamma$-ray fluxes from  other pulsar-driven nebulae, based  
on the  {\it observed}  synchrotron X-ray fluxes  as well as   on the  {\it model assumptions} 
concerning  the nebular magnetic fields, contain significant uncertainties.
Nevertheless, we believe that the future low-energy IACT arrays 
should be  able to reveal and allow study of the spectral  and spatial
characteristics of  this component of radiation  from the synchrotron
nebula surrounding the Vela pulsar,  and hopefully also 
 from some other pulsars  with  ``spin-down  fluxes''
$\dot{E}/4 \pi d^2 \geq 10^{36} \, \rm erg/s \ kpc^2$ (provided that 
 a noticeable part of the spin-down luminosity can be transformed 
 eventually into the shock-accelerated  TeV electrons).

\section{Performance of 5@5}

The concept of  5@5  is  basically motivated by the necessity of sensitive
``gamma-ray timing explorer'' for detailed studies of
 the characteristics of   transient GeV $\gamma$-ray   phenomena
(Aharonian et al. 2001).  In this regard, 5@5 and GLAST present
two complementary  approaches.  While GLAST with  its almost $2
\pi$ steradian field of view  will allow an effective monitoring
of large number of persistent (quasi-stable)  $\gamma$-ray
sources, 5@5  can provide a deeper  probe of individual highly
variable or transient sources.  For example, 5@5 can detect any
multi-GeV \gr flare  with an {\em apparent luminosity}\footnote{In the case
of sources with strong beaming (like pulsars)  within a solid
angle $\Omega \leq 1 \ \rm ster$,  or sources
moving  relativistically towards the observer (like blazars) with Doppler factor
$\delta_{\rm j} \gg 1$, the {\it intrinsic} $\gamma$-ray
luminosities are smaller than the {\it apparent} luminosities by
factors $\Omega/4 \pi$ and $\delta_{\rm j}^{-4}$, respectively.}
of about  $10^{44} (d/1 \, \rm Gpc)^2 \, \rm erg/s$ at $E \leq 10 \ \rm
GeV$ ($d$ is the distance  to the source normalized to 1~Gpc),  
lasting only a few hours. Therefore the prime  objective 
of instruments like 5@5  will be studies  of  
high energy transient phenomena like  GeV/TeV  flares
of   blazars (and perhaps also   microquasars) as well as  
solitary gamma-ray events like GRBs (Aharonian et al. 2001). 
The second important objective of 5@5 concerns the  
detailed $\gamma$-ray spectroscopy of persistent 
$\gamma$-ray sources in the energy region between 10 GeV 
and 100 GeV.

In the original paper  on the  concept  ``5@5''  (Aharonian et al. 2001), 
the following configuration for the telescope array  has been proposed.  
The system consisting  of 5 IACTs is to be installed  at the 
altitude $H=5 \, \rm km$ a.s.l., close to the ALMA (the Atacama Large Millimeter 
Array) site  in Northern Chile. Four telescopes are located at the
corners, and one in the center of a square with a linear size
$d=100 \, \rm m$;  each telescope has an approximately 20-30 m
diameter optical reflector, and is equipped with a multi-channel
conventional PMT based camera with a pixel size $0.12^{\circ}$ and
an  effective field of view FoV$=3.2^{\circ}$. 

\subsection{Detection area, angular and energy resolutions}

To understand  the potential of 5@5 
we need two basic  parameters -- the  effective detection area
$A_{\rm eff}(E)$ and  the angular resolution
$\phi(E)$ as functions of photon energy, approximated
in the range  from $\approx 1.5$ GeV to 100 GeV  in
the following convenient analytical forms:
\begin{equation}
A_{\rm eff}=8.5 E^{5.2} [1+(E/5 \, \rm GeV)^{4.7}]^{-1} \, \rm m^2 \, ,
\end{equation}
and
\begin{equation}
\phi=0.42 (E/5 \, \rm GeV)^{-0.4} \, \rm degree \, ,
\label{angle}
\end{equation}
where E is the energy of the primary  $\gamma$-ray photon
or electron (Aharonian  et al. 2001).
The detection area has a strong energy-dependence  below 10 GeV,
$A_{\rm eff}(E) \propto E^{5.2}$, but at higher energies
 the detection area increases slowly with energy,
$A_{\rm eff}(E) \propto E^{0.5}$. Close to 1 TeV  the detection area
actually becomes constant, in essence
because of the  limited field of the view
of the camera.  The parameter  $\phi$, which  corresponds to
the half-angle  of the cone centered on  the source  and
containing $67\%$ of events at energy $E$,  defines the energy
dependence of the Point Spread Function (PSF) of the telescope.

The energy resolution of the instrument depends
on the  energy; at energies above 10 GeV it could be close to
15-20  per cent.  In this paper for simplicity we   assume an average  
estimate for the  energy resolution, namely a constant value at 
the level of 20 per cent  throughout the entire  energy interval.

The above numbers  characterizing  the basic parameters  of 5@5 
are based on the preliminary Monte Carlo calculations 
(Aharonian et al. 2001) and need to be 
confirmed  by independent simulations, especially   
in the  sensitive  energy region of primary $\gamma$-rays  below 
10-20 GeV  where the  atmospheric cascades consist of only  
a few generations of secondary particles.   
 Further note that the assumed configuration of telescopes 
 is based on  our  ``best guess''. The 
optimal  configuration of both the individual telescopes  
(the area of the dish and  the field of view and pixel size of the imager) 
and  the system (the number of units and average distance between the telescopes) is  a subject for
 future comprehensive  Monte Carlo studies. One cannot exclude that 
the new performance studies  may  result in  some  deviations  from the parameters used in 
this paper.   However, we believe  that the possible  uncertainties in  the 
flux sensitivities should not  exceed a factor of 2 or 3.  Therefore, the main conclusions 
of this paper concerning  the potential of the ``5@5'' concept  for pulsar studies 
are rather  independent of specific configurations of future sub-10 GeV 
IACT arrays.

%
\begin{figure}
\centerline{\psfig{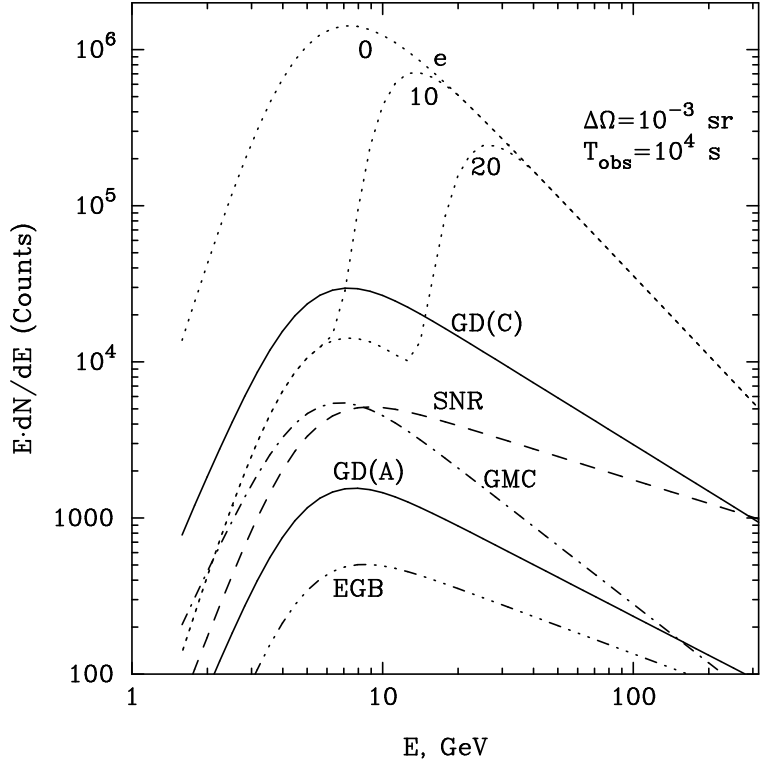}}
\caption{The energy distribution of  background events 
accumulated during observation time
$T_{obs}=10^4 \, \rm s$  within  the solid angle $\Delta\Omega=10^{-3} sr$, 
contributed from the following channels: (1) by  
cosmic-ray electrons  (dotted curves  curves marked as 
``0'', ``10'', and ``20''  correspond  to geomagnetic cutoffs at 
0,  10, and 20 GeV, respectively); (2) from the diffuse  galactic $\gamma$-ray
background  in the direction  of the galactic  center (GD(C)) and
anticenter (GD(A));  (3) from the diffuse  extragalactic  $\gamma$-ray
background  (EGB);  from (4) a ``typical'' molecular cloud
(GMC), and (5)   a ``typical''  supernova remnant
(SNR) (see the text for details). }
\label{fon}
\end{figure}

\subsection{Background conditions}
At energies  below  30   GeV 
the  background  detected by the  system of IACTs
operating in the stereoscopic mode is  dominated by
electromagnetic  showers produced by cosmic ray electrons.
The dominance of the electronic component
is explained by the combination of several factors, in particular
(i) by the large, up to a factor of 5 to 10   difference between
the energies of electrons and protons producing the same amount of
Cherenkov light,  (ii)  by the high altitude of observations,
(iii)  by compact (compared with hadronic showers)
Cherenkov images of electromagnetic showers,
(iv)  by significant   increase (approximately  $\propto E^{-0.5}$)
of the electron-to-proton ratio  of cosmic rays down to
$E \sim 10 \, \rm GeV$. This  offers a unique  method  for the  continuous
({\it on-line}) calibration and control of the detector characteristics
using the  cosmic ray electrons (Aharonian et al. 2001).
The differential spectrum of cosmic ray electrons in the 
region from several GeV to 1 TeV can be approximated as:
\begin{equation}
{dJ_{\rm e}(E)\over d\Omega}=1.36 \times 10^{-3} \
E^{-1} [1+(E/5 \, \rm GeV)^{2.2}]^{-1} \, \rm  cm^{-2} s^{-1}  sr^{-1}  GeV^{-1}  .
\label{elctr}
\end{equation}
The experimental data used for this approximation
contain large, up to factor of 2   uncertainties,   not only at very
high energies, but also in the most critical  energy region
around  10 GeV.  Moreover,  in this energy region  the
flux of electrons entering the Earth's atmosphere
depends on the geomagnetic field, and therefore
on the detector location. For the  ALMA site
with coordinates  $23\degs$ S and $67\degs$ W, the geomagnetic
cutoff of the electrons in the vertical direction
is about  13 GeV (Quenby \& Wenk 1962). This value
agrees with the analytical prediction for the dipole
magnetic field with the equator shifted to the south to
$10 \degs$  at  this longitude (Adams et al. 1981).  The estimates in
the dipole  approximation  give geomagnetic cutoffs  in the  electron spectra
close to the horizon  in  the  eastern and western directions 
 at 9.4 GeV and 33 GeV,
respectively.

In Fig.~1 we show the detection rates of primary electrons by 5@5 (dotted
curves) calculated for 3 different assumptions concerning the
geomagnetic cutoff -  0,  10 GeV, and 20 GeV.  For the
ALMA site the geomagnetic  cutoff  always  exceeds 10 GeV; 
for a  broad range of  optimal (for Cherenkov observations) 
zenith angles   $\leq 40 \degs$,   the geomagnetic cutoff varies between 
10 and 20 GeV.   Since the electron spectrum below
the cutoff has a  very sharp (almost truncated) form, the cosmic
ray background in this energy region  is dominated by protons of 
energy above the geomagnetic cutoff.  It should be noted, however,  that 
the   geomagnetic  cutoff is relevant only to the primary electrons, 
but not to the secondary ones  produced in the atmosphere by effects of
 high energy protons. Therefore  we should expect 
electron fluxes extending well below the geomagnetic cutoff. 
The recent measurements  of this component of electrons
at energies below 10 GeV  by the Alpha Magnetic Spectrometer (AMS) indeed revealed 
non-negligible fluxes comparable with the flux of  primary electrons (Alcaraz et al. 2000).  
Thus this  component   cannot be  {\em a priori}  neglected  in  estimations  of potential 
contributors to the background for space and ground based gamma-ray 
instruments. However, for 5@5 type of instruments the
contribution  of this component fortunately appears  to be quite modest, namely 
less than the  contribution from hadronic showers   induced by high energy 
(above the  geomagnetic cutoff)  primary protons,  
but  accepted by the telescopes at both the hardware  
(trigger) and software  (image identification) levels as low   (sub geomagnetic cutoff)
energy  electromagnetic  showers.  The reason is  that  the trapped  
electrons  spend most of their  time  in  the volume where the  AMS did 
measurements (e.g. Lipari 2002), therefore   the flux of secondary electrons  close 
to the Earth's atmosphere  is estimated   significantly (by an order of magnitude)  
lower  than the fluxes observed at the height  of the  AMS orbit (Lipari, private communication).

At an observation altitude of 5 km
a.s.l.  the  30-100 GeV protons produce
approximately the same amount of Cherenkov light than the
$\gamma$-rays and electrons with energy $\approx 10 \ \rm GeV$.
Therefore, the cosmic ray protons with energy above the
geomagnetic cutoff  initiate  showers a
non-negligible  part of which could be accepted as low energy
electromagnetic showers.   The detailed study
of such ``small-proton'' events requires  detailed Monte Carlo
simulations. Our preliminary study  shows  that  
the integral rate of such events  does not
exceed  10 per cent of the detection rate of cosmic ray electrons.  
Moreover, the further analysis of the shapes of these events should allow
an additional (by a factor of 3 or so) suppression of the proton
background. Therefore  in Fig.~1 we assume
that below the rigidity cutoff the cosmic ray background rate does
not disappear but stays at the level of several  per cent of the
``nominal'' (i.e. without rigidity cutoff) electron detection
rate.  Such an approximation should be confirmed by  future
detailed studies of the  {\em misidentified}  low-energy hadronic showers.  
Therefore,   the detection rates shown in Fig.~1 by the curve {\it without} 
geomagnetic cutoff  and by two curves  {\it with} geomagnetic 
cutoff at 10  and  20 GeV should 
be considered as  conservative upper and lower limits for the 
cosmic ray background.

The isotropic diffuse extragalactic background (EGB) measured
by EGRET in the energy interval 5-300 GeV is approximated as
(Sreekumar et al. 1998)
\begin{equation}
{dJ_{\rm EGB}(E)\over d\Omega}=
3.6\cdot 10^{-8} (E/{5 \ \rm GeV})^{-2.15} \ \rm ph/cm^2 s \ sr \ GeV \ .
\label{iso}
\end{equation}

The fluxes of diffuse galactic  $\gamma$-radiation,  produced
by interactions of cosmic rays with the interstellar medium,
and partly contributed by unresolved galactic sources, strongly
depend on  the galactic coordinates. The largest flux of this
component  arrives  from the  direction of the  galactic center.
For example,  the average flux  from the region 
$l= 0^{\circ}$ and $2^{\circ} \leq |b| \leq 6^{\circ}$
can be approximated as   (Hunter et al. 1997)
\begin{equation}
{dJ_{\rm GC}(E)\over d\Omega}
\approx 3 \cdot 10^{-6}  (E/5 \ {\rm GeV})^{-2.5} \,
\rm ph/cm^2 s \ sr \ GeV \ .
\label{GC}
\end{equation}
The diffuse background from the  region
$l=180^{\circ}, 6^{\circ} < b < 10^{\circ}$,
i.e. in the direction of the Anticenter,
is approximated as (Hunter et al. 1997)
\begin{equation}
{dJ_{\rm GA}(E)\over d\Omega}=
1.45\cdot 10^{-7}(E/5 ~{\rm GeV})^{-2.33} \,
\rm ph/cm^2 s \ sr \ GeV \ .
\label{GA}
\end{equation}

In the case of location of an  weak $\gamma$-ray pulsar  close to
(or behind)  a relatively young supernova remnant  (potential sites of cosmic ray
accelerators and, therefore,  potential high energy $\gamma$-ray emitters),
or  giant molecular  clouds  (radiating  $\gamma$-rays  due to
interactions with the ``sea'' of  galactic cosmic rays) additional
contamination from these extended sources (with angular radius
up to  $1^{\circ}$ or so) should be taken into account
(see e.g. Aharonian 1995).  For a ``typical''  $\gamma$-ray
emitting SNR,  we   use the following approximations
\begin{equation}
J_{SNR}(E)=4 \cdot 10^{-9}
(E/{5 \ \rm GeV})^{-2} \  A_1 \ {\rm ph/cm^2s \ GeV} \, \,
\label{SNR}
\end{equation}
where $A_1=W_{50} n_1/d_{\rm kpc}^2$;
$W_{50}=W_{\rm p}/10^{50} \, \rm erg$ is the total energy of
accelerated protons contained
in the remnant, $n_1=n/1 \, \rm cm^{-3}$ is 
the density of  ambient gas, and $d_{\rm kpc}=d/1 \, \rm kpc$
is the distance to the source.

For a ``typical'' GMC  we adopt the flux
\begin{equation}
J_{MC}(E)=5.6 \cdot 10^{-10} (E/ 5 \ {\rm GeV})^{-2.75} \ A_2
\  \rm ph/cm^2 s \ GeV \, ,
\label{MC}
\end{equation}
where $A_2=M_5/d_{\rm kpc}^2$; $M_5=M/10^5 M_\odot$
is the the mass of the molecular cloud.

The detection rates of  cosmic ray electrons, as well as of the diffuse 
galactic and extragalactic $\gamma$-ray backgrounds within   
$\Delta\Omega=10^{-3} \ \rm sr$ solid angle (opening angle $\approx 1^{\circ}$) 
are calculated with
 \begin{equation}
R(E)=\int  J(E')A_{eff}(E')\Delta\Omega M(E-E')dE' \, ,
\label{fons}
\end{equation}
where  $M(E-E')=\exp((E-E')^2/2\sigma^2)/(\sqrt{2\pi}\sigma)$ is
the energy spread function  for  the Gaussian
distribution with  $\sigma = E/5$ (energy resolution $20\%$).

For the SNR and GMC type of sources
with ``typical'' values for the scaling parameters
$A_1=0.1$ and $A_2=1$, respectively  (Aharonian 1995),
and  assuming $\sim 1^{\circ}$  angular radius
(solid angle $\Delta\Omega = 10^{-3}$),
\begin{equation}
R(E)=\int J(E')A_{eff}(E')M(E-E')dE' \, .
\label{sources}
\end{equation}

In Fig. \ref{fon} we show the differential energy distributions 
of detected background events.   The corresponding total number 
detected  counts  accumulated during the exposure  time 
$T_{\rm obs}=10^4 ~\rm s$ (typical observational time of a single
source by  Cherenkov telescopes  during one night)  
in the energy  range 3  to  300 GeV are: (i) from cosmic rays --
$8.1 \cdot 10^5$  and  $2.8\cdot 10^5$  for 
geomagnetic   cutoffs  at 10 GeV  and 20 GeV, respectively;
(ii) $5.6\cdot 10^4$  and $3.2\cdot10^3$ counts for the
diffuse $\gamma$-ray  galactic emission from  directions
to  the Galactic Center  and Anticenter, respectively;
(iii) $1.2 \cdot 10^3$ counts  for the diffuse
extragalactic background; (iv)    $1.3\cdot 10^4$ and
$9.4\cdot 10^3$ counts   from  ``typical'' SNRs and GMCs.
These results show that except  for the most  ``contaminated'' 
directions towards the   Galactic  Center, the  cosmic ray induced 
background  always dominates over the  contributions from the 
extragalactic and  galactic  $\gamma$-ray backgrounds.

%
\begin{figure}
\centerline{\psfig{file=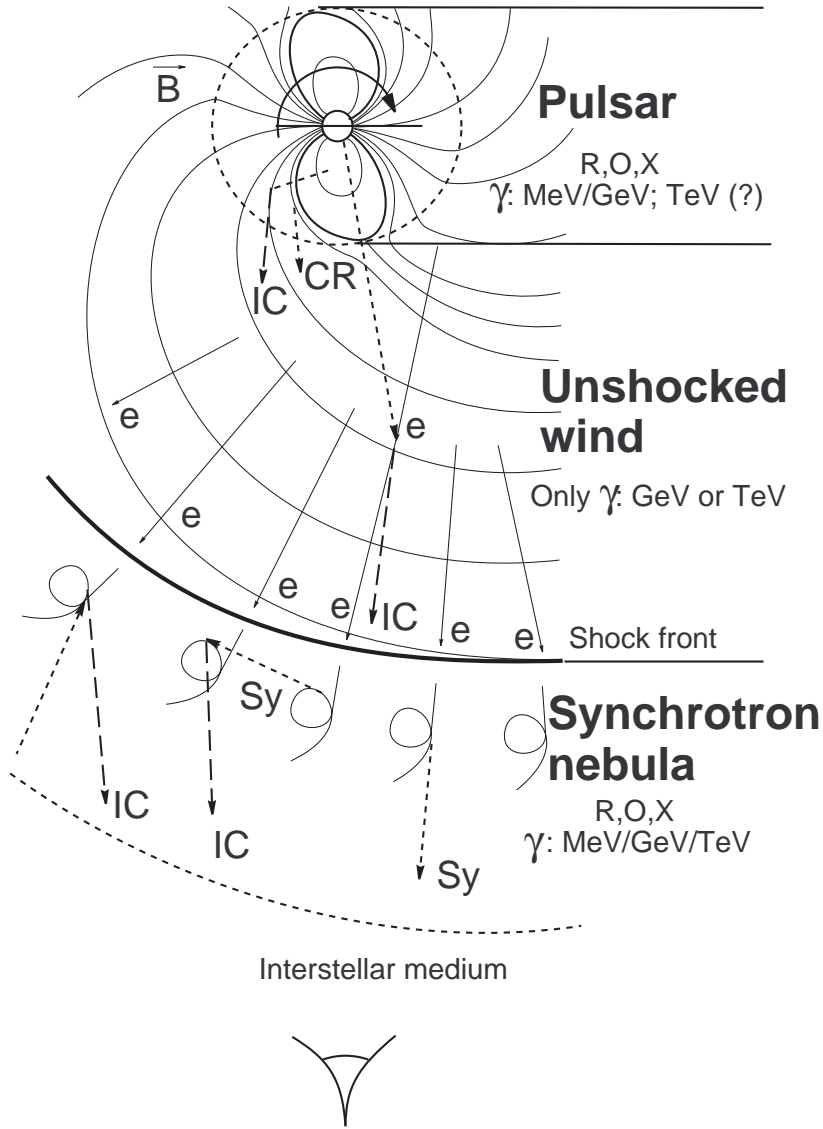,width=13.0truecm,angle=0}}
\caption{\small \label{f1}
Sketch of the sites and radiation mechanisms  
of nonthermal emission  associated with  rotation 
powered  pulsars:  (i) the region  within the  light cylinder 
where the  magnetospheric pulsed radiation  from radio to  
$\gamma$-rays is  produced; 
(ii)  the part of the wind  of cold relativistic plasma close to the  light cylinder
which  effectively emits  GeV  and  TeV $\gamma$-rays through the
IC  mechanism;   (iii) the surrounding synchrotron nebula (plerion)  
which emits  broad-band electromagnetic radiation from radio  to  
multi-TeV   \grs  through the synchrotron and  IC  channels.}
\end{figure}

\section{High energy gamma-rays from RPPs}

High energy \grs  from RPPs  are produced in 3 physically distinguished regions 
through several radiation mechanisms.
 These sites  are schematically shown  in 
Fig. \ref{f1}. The pulsed \grs are generated in the pulsar magnetosphere limited by the 
light cylinder. The relativistic wind ejected by the pulsar can itself produce \grs as well 
as  to excite  the surrounding nebula  which emit electromagnetic emission in a wide 
electromagnetic band.  Gamma-rays  from these regions carry  crucial information about 
the pulsar  and  its interaction  with the surrounding medium. 

The brightness temperature of the pulsed radio emission  is of the 
order of $10^{23}- 10^{30}$~K (Manchester \& Taylor 1977)
implies that  this emission is generated coherently by beams
of relativistic electrons.  In the IR/optical and X-ray bands
the emission is not coherent; it  may have thermal  or nonthermal 
origin. The specific mechanism(s) responsible  for the radio
(e.g. Lyubarsky 1995) as well as  IR/optical and X-ray  (Pacini 1971, 
Becker \& Tr\"umper 1997, Yadigaroglu \& Romani 1995,  
Harding  \& Zhang 2001) remain highly uncertain even for the most 
prominent objects  -- the  Crab and Vela pulsars.

The situation is somewhat better  in the $\gamma$-ray band.
In general,  three   $\gamma$-ray  production mechanisms --
the curvature radiation, synchrotron  radiation,
and inverse Compton scattering --  contribute to  the 
magnetospheric $\gamma$-radiation from
MeV to TeV energies.  Also,  it is recognized that the
pair cascade processes  in the magnetosphere,  play an
important role in the formation of broad-band \gr spectra
(Daugherty \& Harding 1982).  One of the key
questions in  the theory of \gr pulsars remains  the  site(s)
of \gr production.    Presently  two  type  of models  are
discussed in this regard - the    polar cap (Arons 1981,
Daugherty \& Harding 1982,  Usov \& Melrose 1995)
and outer gap (Cheng et al. 1986, Romani 1996,
Hirotani \& Shibata  1999) models.
In both models the primary \grs are
produced  by  high energy electrons due to the curvature radiation.
Some fraction of  the curvature photons are converted into the
$e^{\pm}$ pairs which initiate electromagnetic
cascade in the magnetosphere.  

The winds of $e^{\pm}$ plasma ejected by RPP carry off the major
fraction of their rotational energy.  While initially the wind is dominated by the 
Poynting flux,  at the termination shock
almost all energy  is believed to be  in the form of  kinetic energy  of wind's
bulk motion (e.g. Kennel and Coroniti 1984).  The mechanism which could 
provide such an efficient transformation of the rotational energy  
into the kinetic energy of the cold pair wind is unknown.  The identification 
of this mechanism remains a challenge  for the pulsar physics
(Michel 1982, Coroniti 1990, Lyubarskii \& Kirk 2000).  

The bulk Lorentz factor of the unshocked wind is believed to be   
within $10^4 - 10^7$.  Although the wind is magnetized,  it does 
not emit synchrotron radiation because the electrons  of the wind 
move together with the frozen into the plasma magnetic field. At 
the same time such a wind could be directly observed through its 
Inverse Compton (IC) radiation. In fact, the  inverse Compton 
emission of  the wind is unavoidable because of  bulk motion 
Comptonization by external  low-energy photons of different origin.  
It has been  shown that the illumination of the wind's base  by soft 
thermal photons from the surface of  the neutron star and  by 
nonthermal radiation from the  magnetosphere  should be     
accompanied by  noticeable  IC \gr emission  (Bogovalov
\& Aharonian 2000). The  IC photons  are expected in the energy
interval between 10 GeV  and  10 TeV,  depending on the wind's Lorentz factor.

The  pulsar winds  terminated in the  interstellar medium result in  
strong  shocks  which lead  to the formation of  
synchrotron and IC nebulae around the pulsars. 
The broad-band nonthermal emission  of the Crab Nebula 
(De Jager \& Harding 1992, Atoyan \& Aharonian 1996),
calculated within the  model of  Kennel \& Coroniti (1984), 
is in general    agreement with  observations. Not that 
while  the spectrum of IC radiation of the unshocked wind
is determined primarily  by the wind's Lorentz factor,
the latter has less (direct) effect on the broad-band spectrum 
of the plerion.   
There are good reasons to believe that faint X-ray synchrotron and IC 
nebulae should  surround many other pulsars as well (Aharonian et al, 1997).  

Below we discuss the expected characteristics of $\gamma$-radiation from 
these three regions  for  the case of   Vela pulsar,  the brightest 
GeV $\gamma$-ray  source on the sky  which  can serve  as a standard candle  
for the  low threshold ground-based instruments in the Southern Hemisphere. 

\subsection{Magnetospheric multi-GeV radiation from Vela}

All pulsar models predict that the pulsed radiation above 5 GeV is
dominated by the  curvature radiation of ultrarelativistic
electrons accelerated in the quasi-static electric fields. It  is
generally believed that the charge density in the pulsar
magnetosphere is high enough to screen the electric field $E_{||}$
parallel to the magnetic field,  thus  the corotation condition
${\bf E\cdot B}=0$ is maintained everywhere except for  a few
locations. The regions where  $E_{||} \ne 0$ can exist  quite close to
the surface of the neutron star -- in the polar caps ({\em polar
cap/inner gaps} model),  or  at a  distance comparable to  the
light cylinder  along the null charge surface defined
by the condition ${\bf \Omega\cdot B}=0$, where the corotation
charge density changes sign ({\it outer gap} model) (see, however, 
Hirotani \& Shibata 2001).

\begin{figure}
\centerline{\psfig{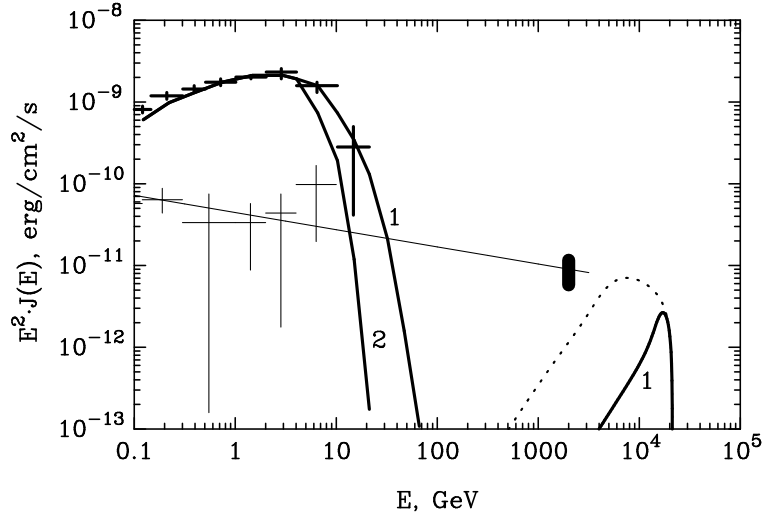}}
\caption{\small \label{f2} 
Gamma-ray  spectra from  Vela.
The pulsed magnetospheric MeV/GeV  emission is shown  by thick solid lines.
The outer gap model (marked as ``1'') predicts exponential cutoff of the pulsed
radiation above 10 GeV.  However, in the TeV range this model predicts
an additional pulsed IC component the intensity of which strongly depends on 
the spectrum  of the magnetospheric IR emission.
Solid curve  in the 10 TeV region corresponds to the IR emission 
with power-law index 1.8 and a  low energy cutoff at 0.01 eV, while the 
dotted line corresponds to the IR cutoff at 0.001 eV.  
The spectrum  predicted in the polar cap model (marked as  ``2'')
has  sharp super exponential cutoff below  10 GeV.  
For comparison the  EGRET measurements  
are also shown -- the points with thick error bars indicate the total 
$\gamma$-ray fluxes,  and  the fluxes of the unpulsed component 
are shown  with thin error bars. The flux of tentative detection
of TeV gamma-rays by the CANGAROO collaboration  (Yoshikoshi et al. 1997) is
also shown. The solid line is the  
extrapolation of the unpulsed EGRET  spectra  to TeV energies (Fierro  et  al.
1998).}
\end{figure}

In the polar cap model  (Arons 1981, Daugherty \& Harding 1982)
acceleration of  primary electrons  occurs   at
distances of an order of the neutron star radius.
These electrons are accelerated up to $\sim 10^{13}$ eV and produce
curvature radiation (CR) with
energy of about  $10 ~\rm GeV$.  Although this radiation initially is
emitted along the magnetic field lines,  
 during the  motion of $\gamma$-rays
in nonuniform magnetic field the angle between the wave
vector of the curvature  photons and the magnetic
filed line becomes quite large, so the  $\gamma$-rays
interacted with the B-field to produce electron-positron pairs.
The optical depth  of the pulsar magnetosphere  drastically 
increases with  $\gamma$-ray energy. In the polar cap  model 
this leads to a very sharp cutoff around 10 GeV.

\begin{figure}
\centerline{\psfig{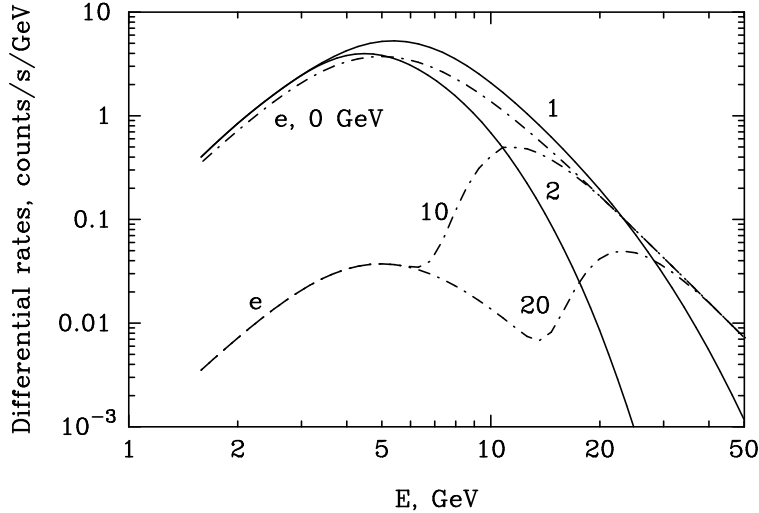}}
\caption{\small \label{rates} Detection rates of magnetosheric 
$\gamma$-rays from Vela by 5@5   expected within the  outer gap  (curve 1)
and polar cap (curve 2)  models. The   background count rates  from
cosmic rays  calculated for 3 geomagnetic cutoff energies at 0, 
10, and 20 GeV are also shown (dot-dashed lines). }
\end{figure}

In the  outer gap model  (Cheng et al. 1986, Romani 1996)
the  electrons are accelerated at larger distances  from the 
neutron star with essentially reduced  magnetic field,
therefore the pair production process  becomes less significant.
The high energy cutoff in the  $\gamma$-ray  spectrum
is defined by the exponential cutoff of the curvature radiation
and by the high energy tail of  the electron spectrum.
For  the truncated electron spectrum, the resulting 
$\gamma$-ray  spectrum has the following  form
\begin{equation}
J(E)={1\over \sqrt{E}}\exp{(-E/E_c)},
\label{cut}
\end{equation}
where $E_c$ is the cutoff energy of the curvature radiation.

The  predicted spectra of pulsed $\gamma$-ray emission by 
the polar cap and outer  gap models (e.g. Thompson 2001) are  shown in 
Fig. \ref{f2}. The absolute fluxes  in both 
cases are normalized to the  fluxes
reported  by EGRET  below several GeV (Kanbach et al. 1994).
The corresponding  differential count rates are  shown in Fig.
\ref{rates}.  The $\gamma$-ray detection rates are  calculated  from
\begin{equation}
R(E)=\int 0.67\cdot J(E')A_{eff}(E') M(E-E')dE' \ ,
\label{eqrates}
\end{equation}
where $M(E-E^\prime)$ is the energy spread function; 
the acceptance factor $0.67$ is due to the
collection of photons  within the angle equal to the 
1~sigma angular resolution.
The background count rates  are  defined as
\begin{equation}
R(E)=\int \Delta\Omega (E)\cdot J(E')A_{eff}(E') M(E-E')dE',
\label{eqrates2}
\end{equation}
where $\Delta\Omega(E)=2\pi(1-\cos{\phi})$.
The calculations of the  integral count rates above $1.8$ GeV yield 
the following values:  $R_{\rm PC}=21 ~\rm s^{-1}$ and 
$R_{\rm OG}=38~\rm s^{-1}$ for the
polar cap and outer gap models, respectively,
the background detection rates are
$R_{\rm e}=5.6$  and  $1.3~\rm s^{-1}$
for the 10 GeV and 20 GeV geomagnetic cutoffs, respectively.

The minimum time for  detection of a  signal at 
5$\sigma$ statistical significance  level  is defined from 
$(N_{on}-N_{off})/ \sqrt{N_{on}+N_{off}} 
= \sqrt{t} (R_\gamma/ \sqrt{2 R_{\rm e}+R_\gamma})=5$, 
where  $N_{on}$  is the number of counts in the direction to the source,
and $N_{off}$   is the counts in the  ``off  direction''.
According to this condition,  a 5$\sigma$ DC-signal   from
the Vela would  be detected by 5@5  after 0.9  sec and 
1.8 sec observations of fluxes predicted by outer gap and
polar cap  models, respectively, assuming 
that the observations are performed at zenith  angles 
corresponding  to approximately 10 GeV geomagnetic cutoff. 
The difference in the total count rates $38-21=17~\rm s^{-1}$
in these  two models  can be measured
on the level of  5$\sigma$ after approximately 3 sec of observation.

The 5@5 type instruments  should  detect  GeV $\gamma$-rays 
from many  other pulsars as well. To  estimate qualitatively the minimum 
detectable fluxes we assume that  the spectrum of Vela is more 
or less typical for other pulsars, and  accept a conservative estimate 
for  the detection rate calculated  for Vela within the polar cap model, 
$R_{\rm PC}=21$. Then,  we arrive at a conclusion that
3~h  observations time should be sufficient  to detect
unpulsed (DC) $\gamma$-ray signals   from pulsars 
at the level of  3  to 10 mVela depending on the available 
observation angles  and corresponding geomagnetic cutoffs. 
The pulsars with energy spectra steeper than the spectrum of Vela
would require more time to achieve such a sensitivity. At the same
time, for pulsars with spectra harder than Vela in the energy region 
above 5 GeV, a deeper probe can be performed for the same observation time.


\begin{figure}
\centerline{\psfig{file=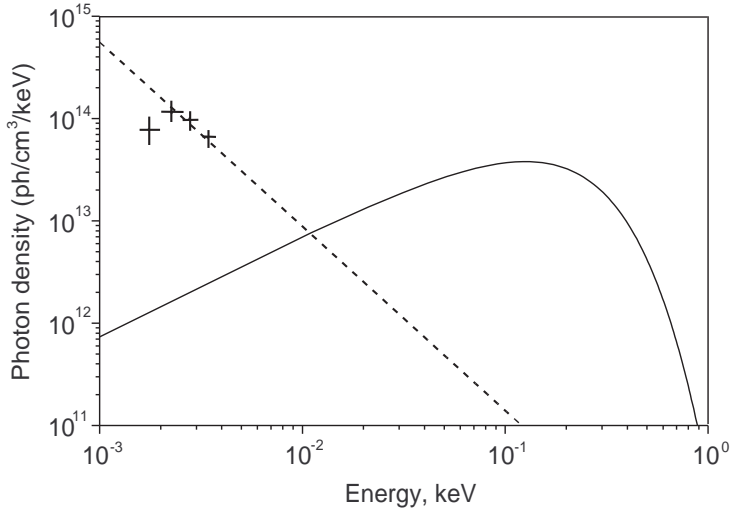,width=10.0truecm,angle=0}}
\caption{Density of soft radiation components at the light cylinder of Vela pulsar.
Observational points  of  optical emission are from  Nasuti et al. (1997). 
The thermal  emission is approximated by blackbody spectrum with
$T=9.2\cdot 10^5~\rm K$ (solid line).  The  nonthermal emission component 
is approximated by a power-law  with  photon index  1.8 (dashed line).
}
\label{dens}
\end{figure}

\subsection{Magnetospheric TeV gamma rays ?}
The high energy \grs produced  in the inner magnetosphere 
are heavily  absorbed before they escape the pulsar, therefore  
 we do not generally expect  magnetospheric TeV \gr emission  
except for  millisecond pulsars that have  relatively modest
 surface magnetic field of about 
 $10^9$ G (see e.g. Bulik, Rudak \& Dyks 2000).  
Within the outer gap models,  non-negligible   TeV $\gamma$-ray
fluxes from pulsars like Vela have been predicted by Romani (1996)  and
Hirotani  (2001).  The basic idea is the following.  The electrons responsible for the
GeV curvature radiation should have Lorentz factor of about $10^7$ or more,
thus in the case of sufficient photon fields these electrons could radiate also 
through the inverse Compton scattering.  
  At first glance, although the synchrotron 
radiation of secondary electrons may serve  
 as a natural  target to support this channel of  $\gamma$-radiation,  
 but it  has been  pointed out by Kwok et al (1991)
that the TeV electrons have a small chance colliding
 with the synchrotron photons because they both move in the same direction. 
Thus,  the production of TeV emission of inverse Compton origin requires other
target photon field(s) which would cross the trajectory of  primary electrons.
Also, the magnetic field in the production region should be rather weak, 
$B \leq 10^9$ G (Usov 1983) in order to avoid the $\gamma$-ray  absorption.
If these two conditions are somehow fulfilled in the GeV $\gamma$-ray 
production region, it is then 
 easy to estimate the expected TeV $\gamma$-ray flux.
 
The outer gap model predicts  that the bulk of \grs  
is produced through the curvature radiation.
 The spectrum of this radiation 
is characterized by a high energy cutoff at  
$E_{\rm c}= 3 \hbar c \gamma^3/2 R_{\rm c}$, 
defined by  the electron Lorentz factor  $\gamma$  and the  curvature
radius $R_{\rm c}$  of the magnetic field line (Ochelkov \& Usov  1980).
Therefore the Lorentz factor of electrons  can be estimated from 
cutoff energy $E_{\rm c}$  in  the observed \gr spectrum,  provided that 
the cutoff does not have other origin, i.e., it
  is the result of \gr  absorption. 
 The intensity of the curvature radiation, 
$\dot E_{\rm CR}=2ce^2 \gamma^4 N_{\rm e} /3R_{\rm c}^2$, 
depends on the Lorentz factor, the total number of accelerated  electrons 
$N_{\rm e}$, and on the curvature radius. 
 Some of the curvature photons can be   converted into secondary 
$e^{\pm}$ pairs which in their turn would emit synchrotron photons  and thus
modify the initial CR spectrum.
 But the full energy flux of $\gamma$-rays,  
 despite of possible spectral modifications,  
 remains close  to $\dot E_{\rm CR}$  because 
 the total energy  of emission is conserved (we neglect here the
energy left with  the cooled secondary electrons). This implies  
$ \Delta \Omega d^2 \int  F(E)dE \simeq \dot E_{\rm CR}$,
where $d$ is the distance to the source and $\Delta \Omega$ 
is the radiation beaming solid angle.  This condition allows us to find  
the parameter $N_{\rm e}$  from the observed \gr luminosity of 
the pulsar with an uncertainty contained in the product $\Delta \Omega d^2$.
Although the combined uncertainty in this parameter could be as large as 
a factor of 10,  it does not have any effect on the estimate of the magnetospheric 
TeV  flux if we normalize  the calculations to the {\em observed} MeV/GeV flux.
  Indeed,  the same electrons responsible for the curvature radiation in the  
MeV/GeV  range,   emit also $\gamma$-rays through the IC channel
boosting  the  soft ambient photons  to TeV energies. The spectrum of this radiation  is 
calculated with 
${\rm d}N/{\rm  d} E_\gamma=\int \sigma(E_\gamma, \gamma) (1-{\bf v c})
n(e)_{\rm ph}cN_{\rm e} de$, 
where  $\sigma(E_\gamma, \gamma)$  is the differential cross section of 
the inverse Compton scattering of an electron with Lorentz factor $\gamma$ and 
velocity  ${\bf v}$ interacting with a photon with energy $e$ and  velocity of motion  
${\bf c}$;   $n_{\rm ph}(e)de$ is the  density of seed photons.  

While the parameter  $N_{\rm e}$ is derived from the observed MeV/GeV $\gamma$-ray energy 
flux, the electron Lorentz factor $\gamma$  is estimated from the observed cutoff in  the
$\gamma$-ray spectrum. Thus, we can write   a simple relation between the resulting 
time-averaged energy flux  of   magnetospheric TeV $\gamma$-rays  of IC origin, 
 ${\rm d}F_{\rm IC} / {\rm d}E_{\gamma}$, and the  observed energy flux 
$F_{\rm CR}$ of $\gamma$-rays   integrated  from  
100 MeV  to  30 GeV, 
\begin{equation}
{dF_{IC}\over dE_{\gamma}}={3\over 2}F_{\rm CR}{E_{\gamma}\over mc^2}({3mc^2\over
E_{\rm c}})^{(4/3)}
({\lambda_{\rm C}\over R_{\rm c}})^{(1/3)}\int \sigma(E_\gamma,\gamma) \ n(e)de \ ,
\label{eq5}
\end{equation}
where  $\lambda_{C}$ is the electron Compton
wavelength. 
Apparently, the light curve 
of the TeV IC  radiation follows exactly  the light curve of the 
MeV/GeV CR  radiation. 
Therefore,  as mentioned above, this  estimate of the IC TeV flux  
does not depend on the actual angular distribution of \gr  emission
and on the distance to the source  if we normalize  
the calculations  to the observed energy flux in MeV/GeV $\gamma$-rays. 

At the same time,  the IC photon flux significantly  depends on the
density and the spectrum of  soft seed photons.
The  densities  of  these photons   at  the light cylinder of Vela are
shown in Fig. \ref{dens}. It consists of two components.  The
thermal emission of the neutron star is described by a
blackbody spectrum.  Several fits were proposed
for this component  (\"Ogelman et al.  1993,  Page et al. 1996, 
Seward et al. 2000).   Here we use the
following parameters $T=9.2 \cdot 10^5 \ \rm K$,
$R_*=10\rm ~km$, assuming  $d=250 \rm~pc$.  
In Fig. \ref{dens} we take into account the possibility for
 the thermal photon density 
 to fall off with distance to the star
 faster than  $r^{-2}$ since the angular distribution
of the photons tends to be mono-directional at the motion from the
star. This effect decreases the photon density at the 
light cylinder by a factor of 2.
At larger distances  the photon density decreases  as $r^{-2}$.
The  power-law (most likely,  nonthermal)  component of soft emission
dominates over the thermal component at optical and 
longer wavelengths.  Optical emission from Vela was measured by Manchester et al. (1978),
\"Ogelman et al. (1989) and Nasuti et al. (1997).
The densities  of  optical photons at the light cylinder  of Vela based on the 
measurements  of Nasuti et al. (1997) are  shown in Fig.~\ref{dens}.
The  recent observations of Koptsevich et al. (2001) with  VLT show that the optical
emission seems to be  smoothly continued into the near IR region down  to 1 eV.  
Unfortunately, no data below 1 eV is available.
The $e^{-1.8}$  power-law approximation to  the observed optical data  
is shown in Fig.~\ref{dens} by the dashed line.

 Fluxes of the magnetospheric TeV radiation,  calculated for two 
different assumptions  for   the position of the low-energy IR cutoff, 
are shown in Fig. \ref{f2}. While the cutoff at 0.01 eV gives 
$\gamma$-ray spectrum with a distinct  line-type feature around 20 TeV,
the infrared cutoff at 0.001 eV makes the spectrum broader with a maximum shifted 
to several TeV. 
 Although the expected fluxes seem to be detectable by 5@5, 
the search for such a pulsed multi-TeV component of magnetospheric radiation  
could be done earlier (and perhaps more effectively)  by the IACT arrays like  
CANGAROO-3 and H.E.S.S. that are better designed 
 for detection of TeV  $\gamma$-rays.  

\subsection{Radiation from the unshocked wind}
The ultrarelativistic winds from the Crab and Vela pulsars are believed to be
essentially cold, therefore the regions of the unshocked wind outflows should be 
under-luminous (no synchrotron radiation). This is well seen in the  Chandra X-ray 
images of the nebulae surrounding the Crab (Weiskopff et al. 2000) and 
Vela (Helfand et al. 2001) pulsars.   However these regions do not  
 have to be  similarly dark  at higher frequencies.  In fact, the unshocked 
ultrarelativistic winds can 
produce non-negligible  high energy $\gamma$-radiation 
via  IC radiation of ultrarelativistic (in the frame of observer) electrons 
moving through the  ambient photon fields. Even in the case of isolated pulsars, 
an adequate photon  density of target photons of both thermal and nonthermal
origins (see Fig.~\ref{dens}) can be  supplied  by the pulsar itself.

The  radiation features of the  unshocked wind in the Crab 
has been discussed  by  Bogovalov \& Aharonian (2000).
A similar  picture should occur in the wind from the Vela pulsar,
although the structures of the winds in these pulsars 
 need not be identical.  The recent observations of the
compact and rather weak nebula  surrounding  the  Vela pulsar
indicate that the  ratio of the  Poynting flux   over the kinetic energy flux  (the
so called $\sigma$ magnetization parameter) is close to 1 (Helfand et al. 2001),
while in the Crab it should be of order  $10^{-3}$ (Kennel and Coroniti 1984).
This parameter describes the ratio of the Poynting flux to the kinetic energy flux of the wind. 
The Poynting flux in the wind is  equal to $\sigma \dot E_{\rm kin}$, 
where $\dot E_{\rm kin}$ is the total kinetic energy flux of the wind, and correspondingly
$\dot E_{\rm kin}=\dot E_{\rm rot}/(1+\sigma)$.
The longitudinal distribution of the wind Lorentz factor, after
the acceleration, follows directly from the energy conservation  law.  
In the simplest case of a wind with isotropic density, 
the Lorentz factor of the wind can be presented as  
$\Gamma(\alpha)= \Gamma_0 +\Gamma_{\rm max}\cos^2\alpha$ (Bogovalov 1999), 
where $\Gamma_0 \approx 100$ is the initial Lorentz factor of the
plasma produced in the pulsar magnetosphere (Daugherty \& Harding,
1982), $\alpha$ is the longitude above the rotational equator of
the pulsar.  The recent Chandra  observations of  Vela  give 
$\alpha= 34.6 \degs$  (Helfand et al. 2001). 
This implies  that even for  the  wind  with  an (quasi) isotropic 
density  the energy flux is concentrated  at the  equatorial plane which explains
(Bogovalov \& Khangoulian  2002)    the observed toroidal structures   
in the  inner  parts of the Crab (Weiskopff et al. 2000) and 
Vela (Helfand et al. 2001) nebulae.

An  important feature of the unshocked wind is that the
electrons  have, under certain conditions,  
nonzero toroidal velocity. 
The energy and angular momentum conservations result in  
the relation  (Bogovalov \& Tsinganos 1999)
\begin{equation}
1-{r\Omega\over c^2} v_{\varphi} \approx {\Gamma_0\over \Gamma} \ .
\end{equation}
Here $\Omega$ is the angular velocity of the
pulsar. It can be  shown that this equation is valid also
for the plasma flow in an arbitrary non-axisymmetric magnetic field.
Therefore, if the plasma is somehow accelerated to a Lorentz
factor $\Gamma \gg \Gamma_0$,  the azimuthal angle of the wind 
is equal to the ratio  $v_{\varphi} / c = R_{\rm L} / r $ (Bogovalov  \& Aharonian 2000),  
whether the wind is kinetic energy dominated or not.
Here $R_{\rm L}$ is the light cylinder, and 
$r$ is the distance to the axis of rotation.  
From this equation  follows that the acceleration
of the wind   in the equatorial plane to 
$\Gamma \gg \Gamma_0$   is possible at 
distances $R_{\rm w}$ exceeding the radius of the 
light cylinder.  Also, this equation  implies that there is always
nonzero angle $\theta$  between the velocity of 
electrons and the radial direction of photons. 
This angle is defined as $\tan \theta =R_{\rm L} (r^2-R_{\rm L}^2)^{-1/2} \cos\alpha$.
Therefore, the IC scattering of  electrons  on the soft thermal and nonthermal emissions
of the neutron star/pulsar   inevitably results in production of  high energy $\gamma$-rays.

The spectral features of the wind emission depend on the
particle flux in the  wind  $\dot N$ 
  and the Lorentz factor as
\begin{equation}
\dot E_{\rm kin}=\dot N (\Gamma_0+{2\over 3}\Gamma_{\rm max})m_{\rm e} c^2,
\label{nrates}
\end{equation}
We may expect  that $\Gamma_{\rm max}$ of the wind from the Vela pulsar
lies in the same  range as in the   Crab pulsar ($10^4-10^7$).  
Here we treat   $\Gamma_{\rm max}$ as  a free parameter, 
and  determine  $\dot N$  through  Eq.~(\ref{nrates}).  
For the Vela pulsar  with $\dot E_{\rm rot}= 7\cdot 10^{36}\rm erg/s$,   
the magnetization parameter was  assumed $\sigma=1$. 

\begin{figure}
\centerline{\psfig{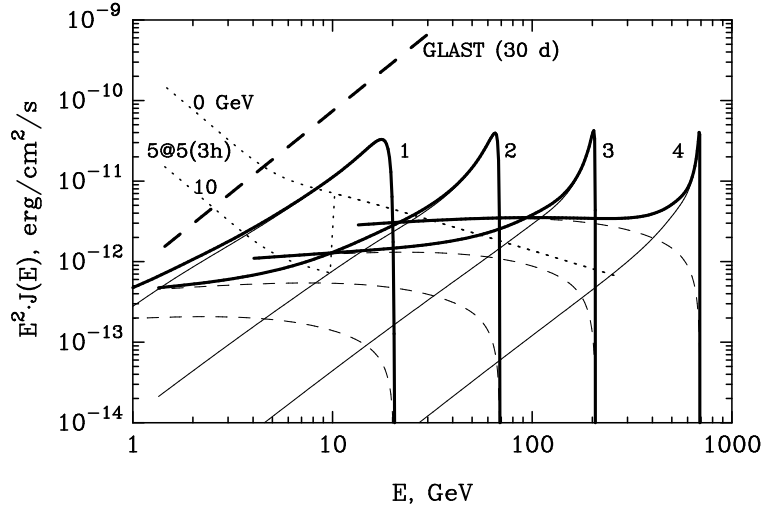}}
\caption{The ``thermal'' (thin solid curves) and ``nonthermal'' (thin
dashed curves)  IC gamma-ray emission components from the unshocked 
wind of Vela. It is assumed that the wind with Lorentz factors 
$\Gamma_{\rm max}=6 \cdot 10^4$ (1), $\Gamma_{\rm max}=2 \cdot 10^5$ (2), 
$\Gamma_{\rm max}=6 \cdot 10^5$ (3) and $\Gamma_{\rm max}=2 \cdot 10^6$ (4)
is accelerated at  $R_{\rm w}=  R_{\rm L} \cos \alpha$. The heavy solid curves 
correspond to the total - ``thermal'' plus ``nonthermal''  $\gamma$-ray emission. 
$3\sigma$ sensitivity threshold of 5@5  for a 3-h observation time is calculated 
for two (0 and 10 GeV) values of the geomagnetic cutoff.
The  sensitivity of GLAST corresponding to 30 d observation time is also shown.
}
\label{wvela}
\end{figure}

In Fig. \ref{wvela} we show the spectra of the IC 
radiation of the wind  for several values of $\Gamma_{\rm max}$.  
The IC emission of the wind consists of two ``thermal'' and ``nonthermal'' 
components produced via  the inverse Compton scattering of the 
wind electrons on the thermal and nonthermal soft photons of the neutron star/pulsar, 
respectively. The spectra of these two components are rather different.
Because of the scattering of the monoenergetic wind electrons on the 
hot thermal photons  with narrow Planckian distribution, 
the spectrum of the  ``thermal'' IC component has  a specific shape with a pile-up
(thin solid curves). The latter is 
especially pronounced  for large Lorentz factors of the wind 
because  the  Compton scattering occurs in deep Klein-Nishina regime.  
The broad spectrum of the nonthermal soft radiation makes
the resulting IC $\gamma$-ray spectrum also very broad 
(thin dashed curves).  Although at low energies this component dominates over the 
``thermal'' IC component,   fortunately  it cannot mask    
the   pile-up of the ``thermal'' IC component that contains direct  and 
unambiguous information about the wind Lorentz factor.  This is  explained 
by the dominance of the thermal radiation over the soft nonthermal 
emission above 10 eV (see Fig.~5).  
Both  $\gamma$-ray components  are generally expected to be modulated
with the pulsar period. While in the case of axially 
uniform wind the  periodic character of the ``thermal'' IC radiation 
could be significantly destroyed, but the ``nonthermal'' IC component  
should be always  
 periodic. This constraint 
 gives  us an additional possibility to separate these 
two components. And finally, it should be noted that the ``nonthermal'' 
IC component can be formed only if the nonthermal soft emission has 
{\em  magnetospheric origin}. Otherwise, i.e. if the production region of the soft 
nonthermal photons coincides  with the wind acceleration region 
(which formally cannot be excluded),  the  ``nonthermal'' IC component would be 
strongly suppressed  because both 
the electrons and photons move in the same direction.       

%
%

\begin{figure}
\centerline{\psfig{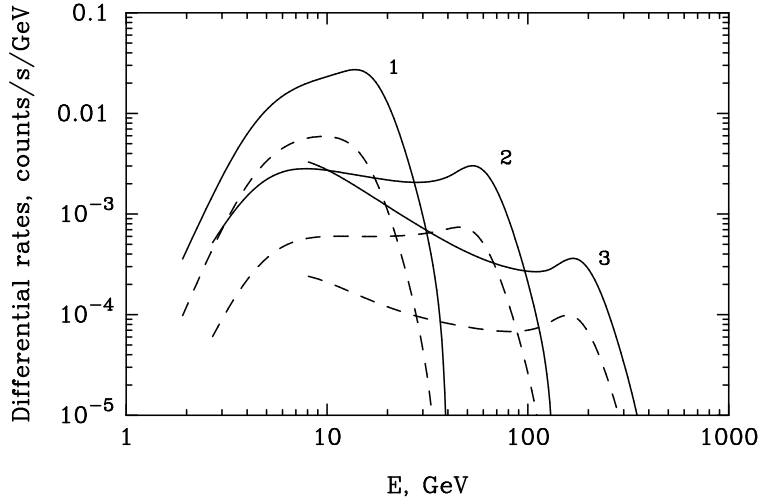}}
\caption{Detection rates of the  total (``thermal''  plus  ``nonthermal'') 
IC $\gamma$-radiation  of the  unshocked wind. The 
Lorentz factors of the wind are the same as in Fig.~6. 
The solid lines correspond to the position of wind   
acceleration at  $R_{\rm w}=R_{\rm L}/\cos\alpha$, 
the dashed lines are calculated for $R_{\rm w}=2R_{\rm L}$.}
\label{examples}
\end{figure}

The $\gamma$-ray fluxes of the unshocked wind shown in 
Fig. \ref{wvela} are calculated assuming that the wind is accelerated 
at the minimum possible distance from the pulsar light cylinder, 
$R_{\rm w}=R_{\rm L}/\cos \alpha$.  We also  show the flux sensitivities of   
5@5  corresponding to  two different assumptions on the geomagnetic cutoff
(0 and 10 GeV). It is seen that in either case the sensitivity and the energy
resolution of 5@5 should be sufficient to detect the  distinct pile-up in the ``thermal'' 
IC radiation spectrum and to discriminate it from  the broad-band emission components 
associated with the ``nonthermal'' IC radiation  of the wind,  as well as from  
the broad-band  smooth IC radiation originating in the  nebula (see Sec.~3.4),
provided that the wind is formed not far from the light cylinder. Unfortunately,
even at this most favorable condition, 
the detection of the radiation of unshocked wind of Vela 
seems not  achievable for GLAST.

In  Fig.~\ref{examples} we show the detection rates of 
IC $\gamma$-rays from the unshocked wind of the Vela pulsar 
calculated for two sites of the wind acceleration 
(more specifically,  in the sites of transition of the wind 
from the Poynting-flux-dominated to 
kinetic-energy-dominated regime) ---  
$R_{\rm w}=R_{\rm L}/ \cos\alpha$ (solid curves)
and   $R_{\rm w}=2R_{\rm L}$ (dashed curves).  
Although the difference in these two values of $R_{\rm w}$ is quite small,
the  difference in expected fluxes and corresponding
detection rates is significant. From results presented in  
Figs.~\ref{wvela} and \ref{examples}
we may conclude that the detection of the IC radiation 
from unshocked wind of the Vela pulsar is possible  only 
if the  wind acceleration takes rather close to the light  cylinder, 
otherwise the  IC \gr flux would be strongly suppressed.
Are there special reasons to believe that the wind is accelerated 
so close to the light cylinder ? We obviously do not know the exact answer
to this question.  However, we may refer the reader to   
Chiueh, Li \& Begelman  (1998) who concluded,  from general analysis of
an ideal MHD flow,  that the acceleration of the wind to a low $\sigma < 1$
state is possible only in the immediate neighborhood of the light
cylinder. Bogovalov (2001)  has argued that the wind
could be accelerated to $\sigma \sim 1$ at the light cylinder 
due to the  magneto-centrifugal mechanism. But  these are theoretical 
arguments. What is most needed  is the 
  direct observational evidence. The results 
in Figs.~\ref{wvela} and \ref{examples} show that 5@5 
should be able to give an answer to  this important question.

\subsection{IC gamma-radiation from the synchrotron nebula}

The termination of  pulsar winds  leads to  
randomization of  velocities of relativistic  electrons
and   formation of bubbles of ultrarelativistic magnetized
plasma  called  pulsar driven  nebulae or plerions. 
The nebulae surrounding the Crab  and Vela pulsars are the most 
prominent representatives of plerions in our Galaxy. 
The wind electrons injected into the Crab Nebula  radiate   almost
$20\%$ of the pulsar's rotational losses  in the form of
synchrotron  X-rays. The size of the X-ray nebula is constrained 
by  the  short lifetime of multi-TeV electrons caused by  their 
synchrotron  cooling. The  nebula in X-rays 
above 0.1 keV has a characteristic toroidal  shape with two jet-like 
features along the  axis of rotation. The linear size of the X-ray toroidal 
structure is about $1.2\cdot 10^{18} ~\rm cm$.
The basic properties of the Crab nebula are well described by 
the hydrodynamical interaction of the cold supersonic pulsar 
wind with the interstellar medium (Kennel \& Coriniti 1984). 

The plerion in Vela consists of a relatively modest  compact nebula
of angular size of 1 arcminute.   For  the distance  to the source
of about 250~pc
this corresponds to a linear size of  $2 \times 10^{17} ~\rm cm$. The
recent Chandra observations revealed  the structure of the compact nebula
which appeared quite similar to the Crab  with the same toroidal shape 
in  the  equatorial region and jet-like features 
along the pulsar's rotation axis. At the same time, 
the  luminosity of the Vela nebula in the 0.1 -10 keV band  is only
$5\cdot 10^{-5}$ of the spin down luminosity of the pulsar. 
The small size and the  low  X-ray luminosity of the compact nebula
was interpreted as indication of a large magnetization parameter
of the  wind in  Vela,  $\sigma \approx 1$ (Helfand et al. 2001),
although alternative interpretations  cannot be excluded.

The fluxes of unpulsed \gr emission reported by EGRET 
in the energy interval between 100 MeV  and 10 GeV 
from Vela   are  shown  in Fig. \ref{f2}.  The  limited angular 
resolution of EGRET (of about $1.5^{\circ}$  at 1  GeV)
does not  allow any  conclusion concerning 
the size of the \gr production region. Because of the 
low X-ray luminosity of the compact nebula, Fierro et al. (1998) 
have suggested  that the unpulsed GeV emission is  produced 
by the pulsar itself. However,   this radiation could be  
produced in a  larger region extended up to 1 degree or so.      
The  Vela pulsar and the compact nebula are embedded in a  hard
X-ray (2.5 -10 keV) structure --- the so-called Vela X. If Vela-X is energized
by the pulsar wind, the pressure in this bubble with volume 
 $V=4\pi R^3/3$ should be $\dot E_{\rm rot}t/V$,
$t=10^4$ yr  being  the age of the pulsar. For the pressure in the surrounding 
remnant of the supernova explosion  $P_{\rm ext} \approx 8.5\cdot 10^{-10} {\rm
erg/cm^{-3}}$, (Markwart  and  \"Ogelman 1997),  the radius of the
bubble is estimated to be $R=8.4\cdot 10^{18} ~\rm cm$ which
corresponds to $0.64\degs$ angular radius, i.e.  quite 
close to the angular size of Vela-X (Willmore et al. 1992).   
Can this  interesting coincidence be interpreted as a hint that  
Vela-X is in fact   a  plerion harboring  the energetic particles 
ejected by the pulsar?
Interestingly, the  power-law extrapolation of the spectrum of 
unpulsed GeV emission (Fierro et al. 1998) towards higher energies 
matches the tentative TeV flux reported by the CANGAROO 
collaboration (see Fig.~3). Given the large uncertainties 
both in the flux and position/size  of the TeV signal, 
we perhaps should not overestimate  the significance such a coincidence.
Moreover, the TeV flux shown in   Fig.~3    is too high 
to be explained by the  most  natural  $\gamma$-radiation mechanism
in the nebula,  by  the inverse Compton scattering on 2.7 K CMBR, unless 
we assume unusually low magnetic field in the production region. Also,
the energy spectrum of \grs predicted by this mechanism is
quite different from a simple power-law  approximation  
in the broad energy region.

\begin{figure}
\centerline{\psfig{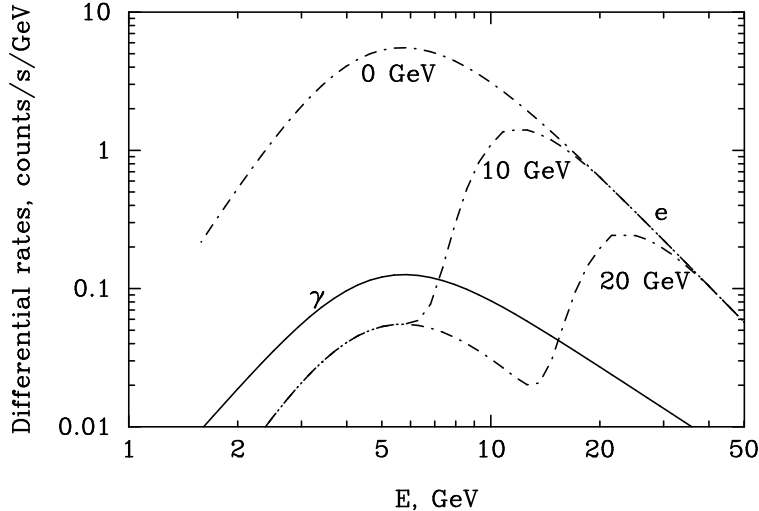}}
\caption{\small \label{neb} Detection rates of the unpulsed 
radiation from Vela (solid line). The  background count rates 
of cosmic rays  within $\theta=0.5\degs$  calculated for 3 different 
values of the geomagnetic cutoff (0, 10, and 20 GeV) are also shown.}
\end{figure}

Nevertheless, 
in order to demonstrate the detection capability of 5@5 it is interesting 
to adopt, as a working hypothesis,  that the radiation of the extended 
region in Vela is described  by a power-law spectrum   of \grs extending  
from GeV to TeV energies as shown in Fig.~3. 
The  expected differential count rates corresponding to this  extrapolation
are  shown in Fig. \ref{neb}.    The integral count rate of the 
unpulsed emission is estimated as $1.7  \ \rm s^{-1}$.
In Fig. \ref{neb} we also show the background rates 
caused by cosmic rays  within the
0.5 \degs angular cone (the angular radius of Vela X) for three  
values of the geomagnetic cutoff
in the electron spectrum. In particular, for 10  GeV geomagnetic cutoff    
the  corresponding integral background count rate
is  $20 \ \rm s^{-1}$ which implies that  the unpulsed
emission from Vela-X can be detected  at the  $5\sigma$ statistical 
significance level  for the  observation  time of about  6 min.

The large \gr photon  statistics and  good angular resolution 
should allow spectroscopic measurements and determination of the size of 
extended plerionic component of $\gamma$-radiation in a broad energy 
region extending from several GeV to  $\sim 1$~TeV. 
Note, however, that such important 
spectrometric and morphological studies above 
100 GeV can be performed  already in the near future by the H.E.S.S. and 
CANGAROO-3 IACT arrays. At lower energies down to 1 GeV, 
 GLAST will contribute key information.

\section{Search of periodical signal from radio-quiet RPP}
The nature of  170 unidentified point-like sources of $> 100$ MeV
\grs  detected by EGRET  (Hartman et al. 1999)  remains mostly unknown.
It is likely that a significant fraction  of these objects 
belongs to  RPPs  with radio beams which do not cross the 
Earth orbit (Grenier 2000).
The beams of \gr emission are believed to be much wider than the
radio beams (see e.g. Romani and Yadogaroglu, 1995; 
Harding and  Zhang, 2001).  Therefore,  a chance to detect
\grs from a pulsar can exceed the chance to detect radio emission.
Obviously  the most direct way to identify the unidentified
EGRET sources with pulsars would be  the detection of 
pulsed  $\gamma$-radiation.

The search  for  periodic signals in the \gr data is not a simple
issue  (see e.g. Chandler et al. 2001).  The
pulsed components of {\it all} EGRET pulsars have been discovered
only after applications of information about the period and
phase obtained at other energy bands. However,  in many cases the
time structure  at different wavelengths could be  different. 
Therefore,  it is crucial to search for periodic \gr
signals without relying on observations at other energy bands. In
 turn, this search requires an adequate \gr photon statistics.
The statistics of photons increases with 
observational time, and in principle could be
very large, in particular for large-field-of-view instruments like GLAST.
However,  the {\it valuable}  observation time can be
increased infinitely only if  the source of the periodic signal is a perfect
clock. Pulsars {\rm are not} perfect clocks. In fact,  
 their periods  {\it increase} with time due to the rotational losses.  
The phase of the signal varies with time as $\varphi= 2\pi(t\nu+t^2\dot \nu)$, 
where $\nu=1/P$ and  $P$ is the period  of rotation, 
$t$ is the time of observations.
The variation of the phase due to the  change of the period of 
rotation  can be neglected while $t^2\dot\nu < 1/2$. Thus, the 
duration of the total time interval for the search of a  periodical 
signal in absence of  an  {\it a priory} information about the period 
and period derivative should satisfy the condition
$T_{\rm obs} < \sqrt{1 / 2\dot\nu}$ \hspace*{0.5ex}.  For example,
the first derivative of the rotational frequency of the Crab  pulsar 
$\dot \nu = 3\cdot 10^{-10}~ \rm s^{-2} $ requires 
observation time 
$T_{\rm obs} \leq 4\cdot 10^4 \ \rm s \approx$ 10~h.
On the other hand,  the number of detected photons
should be large enough to allow determination of the period.
Let's assume that the light curve of the source consists of 
a single narrow pulse. This assumption
agrees  with  the observed pulse shapes  above 5 GeV  for
Crab, Vela, PSR 1706-44 and Geminga pulsars
(Thompson 2001).  In this case  the
epoch folding method allows accumulation  of
$R(>5 \  \rm GeV)T_{obs}$ pulsed photons.  Let's  assume that the
full phase  consists  of $n_0$ bins, and that the  pulse
is  located within one  of these bins.  Then, the m-sigma detection
of the \gr signal in one trial requires
$R_{\rm min} > m \sqrt{R_{\rm b} / n_0 T_{\rm obs}}$,
where $R_{\rm b}$ is the total background   count  rate,
$n_0$ is the number of  bins in the light curve. 
Below  we adopt  5$\sigma$ condition for detection of the
periodical signal. This implies  that the probability to detect
signal by chance multiplied to the number of the trial light
curves should be less then $3\cdot 10^{-7}$. The number of the
trial light curves $N_{\rm tr}=(\nu_{\rm max}-\nu_{\rm
min})/ \delta\nu$, where $\nu_{\rm max}= 10^3 \ \rm Hz$ is the
maximum possible frequency (corresponding to a millisecond pulsar)
and $\nu_{min}=1 \ \rm Hz$ is the minimum possible frequency,
$\delta \nu=1/T_{obs}$ is the frequency separation. Thus, $N_{\rm
tr}\approx 10^7$ for the accepted parameters and therefore the
probability to detect the pulse by chance in one trial should be
less then $3\cdot 10^{-14}$. This means that the $\gamma$-ray
signal  should be detected at the level of  m=7.5.

It is believed that  GLAST will be able to detect as many 
as 250 pulsars,  half of which would be 
unknown at other wavelengths  (McLaughlin \& Cordes 2000).
Also, GLAST can  reveal  the {\em pulsed} emission component 
from a number of  unidentified EGRET sources  if  they  have indeed pulsar origin.
The same can be done by 5@5 type instruments.  Moreover, 
5@5 should be able to identify the  future ``unidentified GLAST'' sources
with pulsars  if the spectra of these objects   extend beyond  5 GeV.  
For  the  background count rate $R_{\rm b}=R_{\rm e}=5.6 \ \rm
s^{-1}$, and assuming $T_{\rm obs}=10^4 \ \rm s \simeq 3 \ \rm h$ (a
standard time of observations of a source  per night) and $n_0=10$,  
for $m=7.5$  we obtain the \gr count rate $R_{\rm min}=0.06 \rm s^{-1}$.  
This corresponds to the flux level of approximately 
3 mVela if we adopt  the minimum possible count rate 
$R \approx 20 s^{-1}$  from Vela as predicted by the polar cap model.

\section{Irregular Variability of the \gr radiation from radio pulsars}

The 5@5 telescope could be quite effective  for studies  of \gr flux variability
from radio pulsars.  Presently at least two types of irregular time  behavior  of radio pulsars are established.
One of them is the so-called ``timing noise" in which the pulse phase
and/or frequency of radio pulses drift  stochastically (Helfand et al. 1980,
 Cordes \& Helfand 1980). Another type of
irregularity  in the radio emission from  radio pulsars  are "glitches"
-- sudden increase in the pulse frequency produced by apparent change in the
momentum of inertia of  neutron stars.
These rather rare events occur
mostly in young pulsars (Reichly \& Downs 1971, Lyne 1996). Glitches were
observed from 27 pulsars
(Wang et al. 2000).

The Vela pulsar seems to be the best object to  study the
irregular variability of pulsars in $\gamma$-rays by 5@5. A dozen
glitches have been recorded over the past three decades from this
source.   EGRET data show that the flux in the range above 100 MeV
is variable.  Remarkably, there is
evidence of variation of the spectral index of the radiation, as
well as  of the relative intensity of two peaks of the light
curve (Ramanamurthy et al. 1995). 
The variability of  \gr emission from the Vela pulsar
detected by EGRET has been derived  for  14-day observations 
 but the photon
flux was too low to investigate the variability on smaller time
scales. 5@5 is expected to allow detection and resolution of
 the $\gamma$-ray variability on sub-hourly timescales.

\section{Summary}

The high flux sensitivity of 5@5,  supported with unprecedented photon statistics
in one of the most informative windows of electromagnetic radiation
for the pulsar physics, $E \geq 5 \ \rm GeV$, should allow detailed
studies of spectral and temporal characteristics of  the
\gr pulsars detected by EGRET,   as well as  very effective searches for
pulsed components from   unidentified EGRET sources.  The
energy and  angular resolutions of 5@5 can provide
unique studies of other radiation components of pulsars 
associated with the unshocked winds and the synchrotron nebulae
(plerions) in a broad energy band extending from several GeV to 1 TeV. We believe  
that the new generation sub-10 GeV ground-based \gr instruments like 5@5
can provide  crucial insight into the physics of pulsars and their interactions
with surrounding interstellar medium.

\section{Acknowledgments}

We are grateful to P. Lipari for his comments concerning the 
interpretation of the  secondary electron  fluxes detected 
at energies below the geomagnetic cutoff by AMS. We thank 
D. Thompson and the second (anonymous) referee for their   
critical remarks  which improve the paper significantly. 
 SB thanks MPI f\"ur Kernphysik for hospitality and support during
his work on this paper. The work was partially supported by
collaborative INTAS-ESA grant N 99-120.

\end{document}